\documentclass[conference]{IEEEtran}
\pdfoutput=1
\usepackage{cite}
\usepackage{amsmath,amssymb,amsfonts}
\usepackage{algpseudocodex}
\usepackage{algorithm}
\usepackage{graphicx}
\usepackage{textcomp}
\usepackage{xcolor}
\usepackage[hyphens]{url}
\usepackage{multicol}
\usepackage{multirow}
\usepackage{circledsteps}
\usepackage{balance}
\usepackage{color, soul}

\def\BibTeX{{\rm B\kern-.05em{\sc i\kern-.025em b}\kern-.08em
    T\kern-.1667em\lower.7ex\hbox{E}\kern-.125emX}}

\newtheorem{definition}{Definition}
\algrenewcommand\textproc{}
\makeatletter
\newcommand\fs@norules{\def\@fs@cfont{\bfseries}\let\@fs@capt\floatc@ruled
  \def\@fs@pre{}%
  \def\@fs@post{}%
  \def\@fs@mid{\kern3pt}%
  \let\@fs@iftopcapt\iftrue}
\makeatother

\floatstyle{norules}
\restylefloat{algorithm}

\title{Dynamic Ineffectuality-based Clustered Architectures}
\author{\IEEEauthorblockN{Rajshekar Kalayappan}
\IEEEauthorblockA{Dept. of Computer Science and Engineering,\\
Indian Institute of Technology Dharwad\\
Dharwad, India\\
Email: rajshekar.k@iitdh.ac.in}
\and
\IEEEauthorblockN{Sandeep Chandran}
\IEEEauthorblockA{Dept. of Computer Science and Engineering,\\
Indian Institute of Technology Palakkad\\
Palakkad, India\\
Email: sandeepchandran@iitpkd.ac.in}
}

\begin{document}
\maketitle
\thispagestyle{plain}
\pagestyle{plain}

\begin{abstract}
The direction of conditional branches is predicted correctly in modern
processors with great accuracy. We find several instructions in the dynamic
instruction stream that contribute only towards computing the condition of these
instructions. Hence, when the predicted direction of conditional branches is
indeed correct, these instructions become {\em Ineffectual} -- the functional
state of the program would not be different had these instructions been dropped.
However, the execution of ineffectual instructions cannot be avoided altogether
because it is possible that the prediction of the branch direction is wrong. In
this work, we determine all sources of ineffectuality in an instruction stream
such as conditional branches, predicated instructions, indirect jumps and
dynamically dead instructions. Then, we propose a technique to steer the
ineffectual instructions away from the primary execution cluster so that
effectual instructions can execute uncontended. We find that such
ineffectuality-based clustering of instructions naturally simplifies the design
and avoids several caveats of a clustered architecture. Finally, we propose a
technique to detect instances when instructions were incorrectly marked as
ineffectual, say due to a branch misprediction, and recover the pipeline. The
empirical evaluation of the proposed changes on the SPEC CPU2017 and GAPBS
benchmarks show performance uplifts of up to $4.9$\% and $10.3$\% on average
respectively.

\end{abstract}

\section{Introduction}
\label{sec:intro}

The microarchitecture of a processor core is popularly envisioned as comprising
two parts -- the front end and the back end. While the front end is
responsible for identifying the work to be done by the core, the back end
performs the actual computations. 

One of the several possible techniques to increase the performance of the core
is to increase the width of the pipeline. This enables the processor to exploit
instruction-level parallelism better.

\begin{figure}
\centering
\includegraphics[width=0.9\columnwidth]{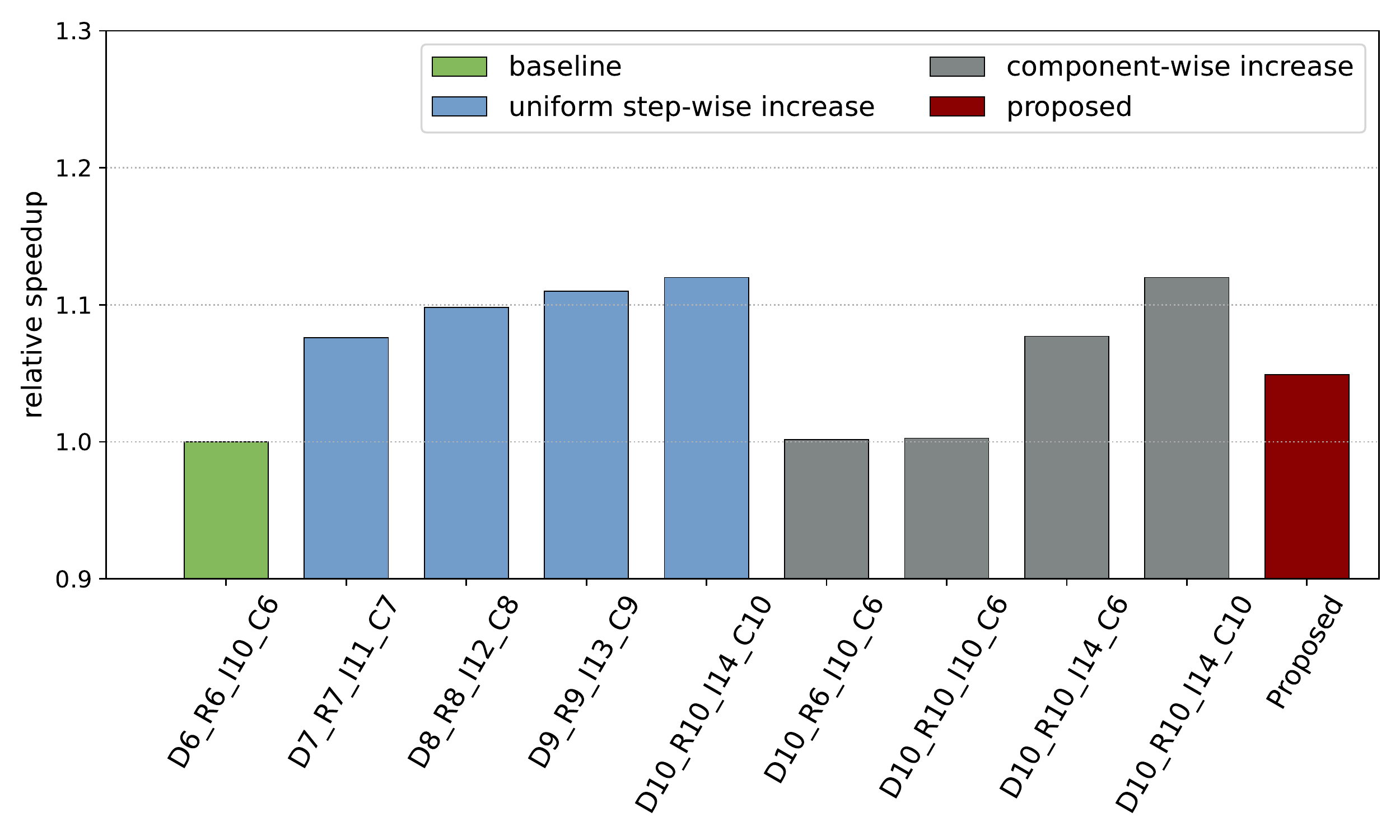}
\caption{Average performance gain seen in SPEC CPU2017 suite with
  wider (hypothetical) pipelines.
  D\texttt{p}\_R\texttt{q}\_I\texttt{r}\_C\texttt{s} denotes a decode width of
  \texttt{p}, rename width of \texttt{q}, issue width of \texttt{r}, and commit
  width of \texttt{s}.}
\label{fig:characterization}
\end{figure}

Figure~\ref{fig:characterization} shows the possible performance improvements,
across the SPEC CPU2017 benchmarks suite, when we increase the width of the
different stages of the pipeline hypothetically. The blue bars indicate a
uniform increase in the width of the pipeline in a step-wise manner (all
components in the pipeline such as decode width, rename width, issue width and
commit width increase by 1 in each step). The grey bars indicate a
component-wise increase in the width of the pipeline, starting from the decode
width. We notice that a wide pipeline ({\em D10\_R10\_I14\_C10}) yields a
significant performance gain ($12$\% on average).

There are four key challenges to increasing the width of the pipeline that have
been identified previously~\cite{revisitingsuperscalarthesis}. The first
challenge is that the front end should be able to fetch and decode enough
instructions to always keep the execution lanes busy. Significant progress has
been made on this front by the incorporation of structures such as micro-op
caches, loop-stream detectors, and instruction prefetchers.

The second challenge is in increasing the width of the result bus that carries the
results from execution units to the register file as well as the reservation
stations. This is typically overcome by grouping the execution lanes into
clusters. The clustered architectures usually partition the result bus. The
implication of such partitioning is that forwarding of results to dependent
instructions within the cluster is fast, whereas inter-cluster communication is
time-consuming and expensive.

The third challenge is to increase the number of read and write ports of the
register file. This is necessary because a wider pipeline needs to read the 
source operands and write the results of a higher number of instructions every
cycle. However, increasing the number of ports of a register file increases its
area quadratically. This issue is typically addressed by partitioning or
replicating the register file in each of the clusters. This allows each
partition to be written by only one cluster but read by all clusters.
This is also known as register-write
specialization~\cite{registerwritespecialization}.

Finally, the fourth challenge is to design a scheduling technique that takes
into account, the higher number of execution lanes. The popular solution is to
use a hierarchical scheduler where a high-level scheduler steers instructions
into clusters and a low-level scheduler (one per cluster) issues instructions
into the execution lanes as and when it is ready. This is non-trivial because
the schedulers would have to ensure that load across the clusters is balanced
while also minimizing inter-cluster communication.

In this paper, we identify another approach that will yield higher performance
as compared to a naive increase in the width of the pipeline. This is done by
identifying instructions dynamically whose execution does not have any impact on
subsequent instructions. We call such instructions, {\em Ineffectual
  Instructions}. Examples of ineffectual instructions are highly predictable
branches as well as the \texttt{cmp} instruction preceding them. We then propose
to use \textbf{asymmetric} (in throughput), but \textbf{homogeneous} (in
execution capabilities), clusters, where a primary high-performance,
out-of-order pipeline executes effectual instructions while a secondary simple,
in-order pipeline executes ineffectual ones. This will allow effectual
instructions to utilize the primary cluster uncontended. The overall performance
of the application is determined by the primary cluster, while the secondary
cluster merely ensures correctness. 

The additional benefits of an effectuality-based partitioning of work between
the clusters are that it naturally (i) eliminates the need for complex bypass
mechanisms between the clusters and time-consuming inter-cluster communications,
as well as (ii) yields itself to effective register-write specialization. Our
studies show that executing ineffectual instructions on a newly added secondary
(but simple) cluster can increase the overall performance of the processor by
$4.9\%$ on average for SPEC CPU2017 as compared to the state-of-the-art
processors.

The rest of the paper is organized as follows. Section~\ref{sec:related}
discusses the related state-of-the-art approaches to improve performance.
Section~\ref{sec:intuition} motivates our solution with an example.
Section~\ref{sec:defs} formally introduces the terms we use to discuss our
proposal. Section~\ref{sec:arch} details our proposed solution, and
Section~\ref{sec:eval} shows the results of our empirical evaluation of the
proposed solution. Section~\ref{sec:conc} has our concluding remarks.

\section{Background and Related Work}
\label{sec:related}

\subsection{State-of-the-art in front end design}
The front end of the pipeline is responsible for identifying the work to be
done. It starts by fetching multiple instructions, or macro-ops, up to 16 bytes
per cycle from the L1 instruction cache. The fetch engine is assisted by a Branch
Prediction Unit (BPU) that predicts the branch targets. State-of-the-art
processors typically achieve prediction accuracies of over $98\%$ (SPEC CPU2017
suite~\cite{cpu2017characterization}). Predecoding is done to handle cases of
Length Changing Prefixes, and the macro-ops are then placed in a macro-op queue.
The next step is to decode or translate macro-ops into RISC-like micro-ops, that
the back end of the pipeline can understand. The decode engine is a complex 
system comprising multiple decoders and provides a peak bandwidth of a few
(5-6) micro-ops per cycle.

The decoded micro-ops are cached, and future references to the same instruction
are served out of the micro-op cache. The hit rate of the micro-op cache is
high and reaches 100\% for certain favorable program phases~\cite{spectre}.
This allows powering down the primary fetch and decode pipelines to save
energy. The micro-op cache functions as a streaming cache, in contrast to
traditional processor caches. This streaming nature allows the micro-op cache to
potentially provide a higher throughput.

\subsection{State-of-the-art in back end design} 

Two types of schedulers are popular in modern processors, {\em Unified}
schedulers and {\em Split} or {\em Clustered} schedulers. In the designs that
use a unified scheduler, all instructions are scheduled onto the FUs by a single
scheduler. Intel microarchitectures typically~\cite{skylake} follow this
approach. Increasing the width of such a pipeline suffers from the challenges
discussed in Section~\ref{sec:intro} and hence, requires significant effort to
achieve high base frequencies.

In the designs that use split schedulers, the FUs are grouped into clusters and 
use multiple schedulers to schedule instructions onto the FUs. AMD's Zen 
series~\cite{zen2} and IBM's Power series~\cite{power10} of processors follow this
approach. The grouping of FUs into clusters helps to reduce the design complexity of  
individual schedulers and helps increase the base frequency of the core but 
introduces additional design challenges. The maximum throughput now depends on two
orthogonal factors~\cite{scalabilityaspectsclustered}: (i) the amount of work that 
each cluster does should be similar (load is balanced), and (ii) the volume of 
inter-cluster communication (consuming values produced in other clusters) should 
be minimal. Inter-cluster communication is required in a clustered architecture 
when instructions fail to obtain the value of a source operand from either 
its local bypass network, or the Physical Register File (PRF). Such inter-cluster
communication is time-consuming and could significantly delay the execution of 
the consuming instruction~\cite{revisitingsuperscalarpaper}. The aggressive 
pursuit of just one of these (say minimizing inter-cluster communication) could
adversely affect the throughput (due to load imbalance).

Several proposals have attempted to balance the two aspects by designing
instruction steering schemes that are cognizant of the dependence between
micro-ops as well as the load on each cluster~\cite{staticclusterpartitioning,
  dynamiccodepartitioning, lowpowerhighperformanceclusters}. A technique to balance 
the load across clusters better is to prioritize critical instructions when the load 
is high on a particular cluster~\cite{fields, criticalitycluster}. An orthogonal
approach is to employ heterogeneous clusters~\cite{intfloatcluster,zen2} -- one
for performing integer operations and the other for floating point and vector
operations. While this reduces inter-cluster communication, it is still
susceptible to workload sensitivity (imbalanced cluster load). In contrast,
since we partition the instructions based on their ineffectuality, the need for
forwarding the results from the secondary cluster to the primary cluster is
eliminated (when the ineffectuality speculation is correct). However, communication
in the opposite direction is still necessary but is not on the critical path, 
and therefore, can tolerate delays. 

The challenge of the area overhead of the PRF increasing quadratically with 
increasing the number of PRF ports (due to an increase in the pipeline width)
is addressed in two ways. The first is to duplicate the PRF to avoid increasing 
the read ports~\cite{kessler1999alpha}. The second is to partition the PRF among 
the clusters, with each cluster allowed to write to only its partition (but can 
read from all partitions) ~\cite{registerwritespecialization} instead of 
increasing the write ports. We employ the same strategies in this work.

\subsection{Dynamic code transformation and elimination}
\begin{figure*}
\begin{center}
\begin{tabular}{cc}
  \begin{minipage}{0.40\textwidth}
  \includegraphics[width=0.9\textwidth]{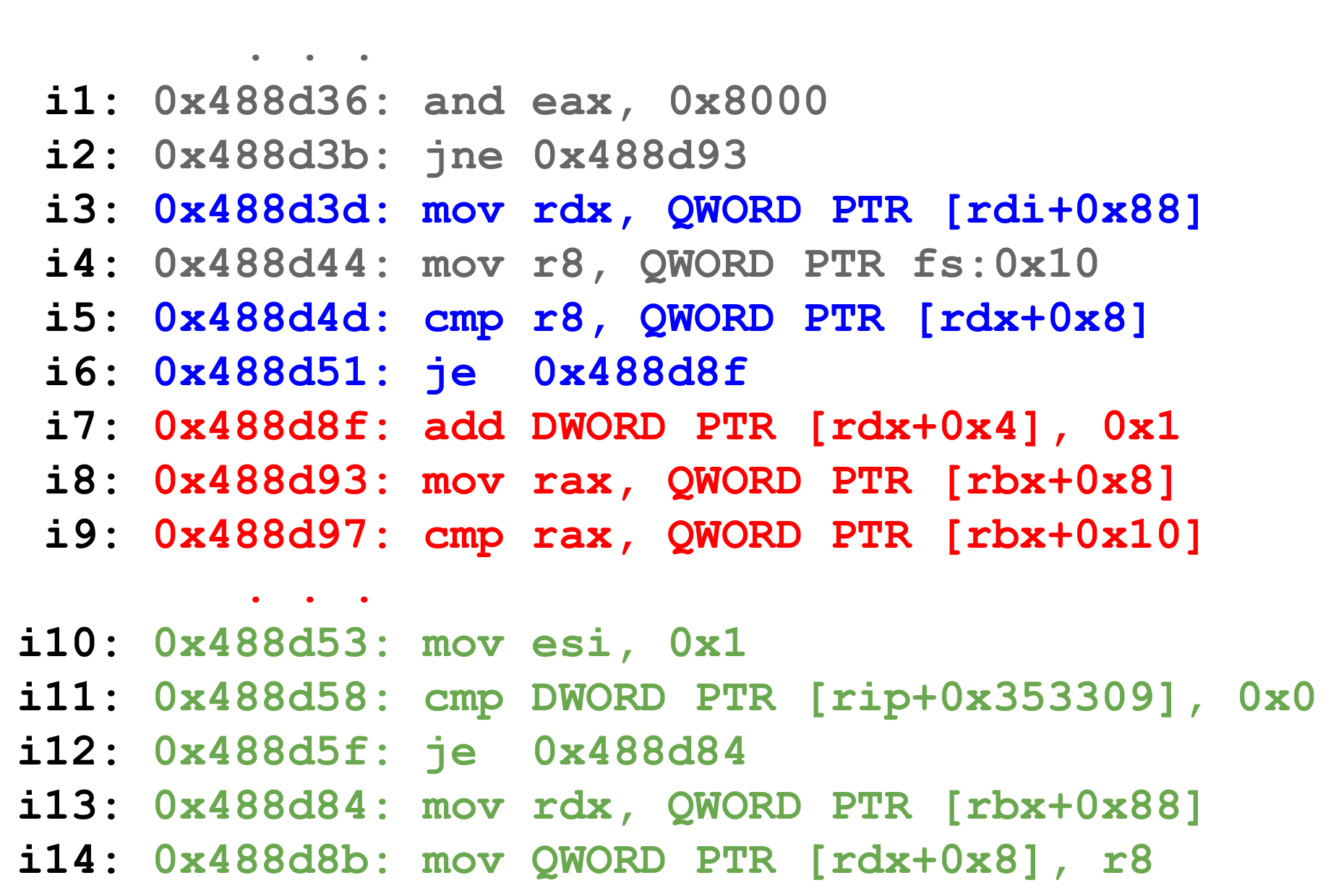}
  \caption{The \texttt{nab} code snippet highlighted to indicate instructions
    that influence the outcome of the conditional branch (indicated in blue),
    the instructions on the wrong path and correct path (indicated in red and
    green respectively).}
  \label{fig:code_execution}
  \end{minipage}

  &
  \begin{minipage}{0.55\textwidth}
  \includegraphics[width=0.99\textwidth]{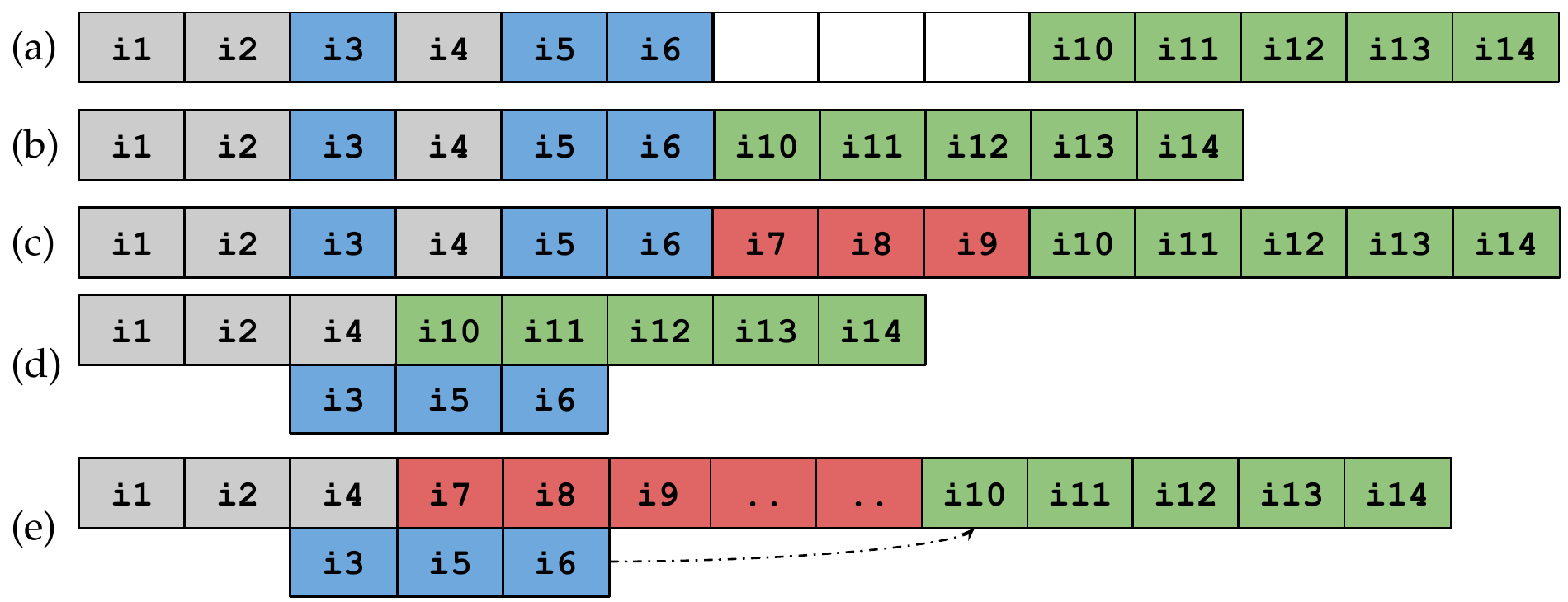}
  \caption{Comparative Analysis of the execution of the code snippet shown in
    Figure~\ref{fig:code_execution} on different architectures: (a) simple
    pipeline without BPU, (b) and (c) show a simple pipeline when $i6$ was
    correctly and incorrectly predicted respectively, (d) and (e) show the
    proposed pipeline when $i6$ was correctly and incorrectly predicted
    respectively.}
  \label{fig:motivation}
  \end{minipage}
\end{tabular}
\end{center}
\end{figure*}

Another approach to improve performance is to drop instructions that are deemed
unimportant by the respective techniques. This helps in freeing up back-end
resources for the remaining instructions, thereby resulting in better
performance. 

The techniques under this approach can be broadly classified as non-speculative
and speculative. The optimization techniques such as move elimination and the
zeroing idiom in the rename stage~\cite{skylake} fall under the
non-speculative category. This category also includes proposals that implement
popular compiler optimizations such as constant folding and common
sub-expression elimination in hardware~\cite{petric2005reno}.

The speculations made by the speculative techniques must be verified before the
instructions are allowed to commit. The techniques such as value prediction,
where instructions with predictable results are removed from the
stream~\cite{petric2005reno, jourdan1998novel, perais2014eole,
  perais2021leveraging} fall under the speculative category. Further, the
overheads of value prediction are reduced by predicting the values of only
critical instructions~\cite{bandishte2020focused}. Another proposal to reduce
the overhead is to save the optimized stream in the micro-op cache for future
use~\cite{moody2022speculative}. Similarly, speculative techniques to predict
memory addresses to identify opportunities for redundant load
elimination, store forwarding, and silent store eliminations have been proposed
in the past~\cite{fahs2005continuous, kim2002implementing}. Another speculative
technique drops dynamically dead instructions from the instruction
stream~\cite{butts2002dynamic}. In a similar vein, our proposed technique also
speculatively predicts ineffectual instructions and drops them from the primary
pipe.

The notion of ineffectuality was introduced in the context of look-ahead
processors~\cite{slipstream}. Here, a stream with reduced instructions
(A-stream), obtained by discarding highly predictable branches and instructions
whose results are never used, is executed ahead of the actual stream of
instructions (R-stream). The A-stream runs ahead and speculatively warms up the
caches as well as passes control flow hints to the R-stream so that the latter
can execute the application faster. The R-stream corrects the A-stream in case
of misspeculations. This approach of using ineffectual instructions is entirely
different from our proposed technique. 

\section{Motivation}
\label{sec:intuition}

Consider the example code snippet from the {\tt nab} benchmark of the SPEC
CPU2017 suite (compiled using GCC with optimization flags enabled) in
Figure~\ref{fig:code_execution} and its corresponding execution timeline shown
in Figure~\ref{fig:motivation}.

Let us now consider the conditional branch instruction $i6$. We look for
instructions whose results \textbf{solely} contribute to the outcome of $i6$. We
see that $i6$ depends on the \texttt{cmp} instruction $i5$, and the latter's
result is not used by any other instruction. Further, $i5$ is dependent on $i3$
and $i4$. The only consumer of the result produced by $i3$ is $i5$, whereas the
result produced by $i4$ is also used by $i14$. Such analysis could continue with
$i3$'s predecessors as well (not shown here). Thus, $\{i3,i5\}$ are the
instructions of interest to us.

Let us further assume that $i6$ is a highly predictable branch. If the branch
outcome is predicted correctly, its execution does not contribute towards the
overall program state. Consequently, the execution of $i5$ and $i3$ also does
not contribute to the overall program state. These instructions are examples of
ineffectual instructions (and marked in blue). However, if the branch outcome of
i6 was predicted incorrectly, skipping these instructions would have led to an
incorrect program state. Therefore, these instructions still need to be executed
to verify the correctness of the branch prediction.

Since the ineffectual instructions do not contribute towards the overall program
state, they can be offloaded from the primary cluster, thereby making way for
effectual instructions (as shown in Figure~\ref{fig:motivation}(d)). Such
offloading of instructions could significantly improve the performance of the
system when the branch outcomes are predicted correctly
(Figure~\ref{fig:motivation}(d) completes ahead of
Figure~\ref{fig:motivation}(b)). Typically, the branch prediction accuracy of
modern processors is over $96\%$~\cite{cpu2017characterization}. We observed
similar performance when we simulated the TAGE-SC-L branch
predictor~\cite{tage_sc_l}. Since correctly predicted conditional branches,
predicated instructions and indirect jump instructions are a source of
ineffectuality, there is a significant opportunity for improving the performance
of the processor. As we will see later in Section~\ref{sec:eval}, the number of
ineffectual instructions in a workload can be as high as $36.4\%$ (see
Figure~\ref{fig:ineffectual}).

Another source of ineffectuality is when data values produced by some
instructions are not used because the control flow did not take a particular
path. For example, consider the instruction $i1$ that writes to the register
\texttt{eax}. If the branch $i6$ is not taken, the result of $i1$
is overwritten by $i8$ before it was ever used. Hence, $i1$ is data ineffectual
(dynamically dead instruction) in this case. Our studies show that
we can identify a significant number of ineffectual instructions when 
we consider data ineffectuality as well.

\section{Definitions}
\label{sec:defs}

We formally state the definitions of the terms we use in the paper. The examples
mentioned here refer to the code snippet in Figure~\ref{fig:code_execution}.

\begin{definition}[Register Dependence]
An instruction $i$ is said to be \textbf{\em Register Dependent} on $j$ if $j$
produces a register value that is then consumed by $i$. For example, instruction
$i5$ is register dependent on $i3$.
\end{definition}

\begin{definition}[Memory Dependence]
An instruction $i$ is said to be \textbf{\em Memory Dependent} on $j$ if $j$
produces a memory value that is then consumed by $i$.
\end{definition}

\begin{definition}[Data Dependence]
An instruction $i$ is said to be \textbf{\em Data Dependent} on $j$ if $i$ is
dependent (either register dependent or memory dependent) on $j$ or is
(recursively) dependent on any of the other dependants of $j$. For example,
instruction $i6$ is data dependent on $i3$, $i4$ and $i5$.
\end{definition}

\begin{definition}[Input Logic Cone]
The \textbf{\em Input Logic Cone} ($ILC_i$) of an instruction $i$ is the set of
instructions upon which $i$ is data dependent. For example, $ILC_{i6} = \{i3,
i4, i5\}$.
\end{definition}

\begin{definition}[Output Logic Cone]
The \textbf{\em Output \\
  Logic Cone} ($OLC_i$) of an instruction $i$ is the set of
instructions that are data dependent on $i$. For example, $OLC_{i4} = \{i5, i6,
i14\}$.
\end{definition}

\begin{definition}[Pivot Instruction]
\label{def:pivot}
An instruction $i$ is called a \textbf{\em Pivot Instruction} if:
\begin{enumerate}
    \item conditional branch that is accurately predicted
    \item predicated instruction whose predicate is accurately predicted as
    \texttt{false}
    \item an indirect jump whose target is accurately predicted
    \item instruction whose $OLC_i$ is accurately predicted to be $\emptyset$
\end{enumerate}
\end{definition}
For example, $i6$ will be a pivot instruction if it is accurately predicted.
$i1$ will be a pivot instruction if $i6$ is not taken.

\begin{definition}[Ineffectual Instruction]
\label{def:ineffectual}
An instruction $i$ is \textbf{\em Ineffectual} if all the instructions in $OLC_i$
are ineffectual. We propose an alternate definition that is amenable for
efficient implementation in hardware.

An instruction $i$ is \textbf{\em Ineffectual} if
\begin{enumerate}
    \item it is a pivot instruction, or
    \item $i \in ILC_j$ where $j$ is ineffectual and all the instructions in $OLC_i$ are ineffectual.
\end{enumerate}
\end{definition}
For example, let us assume that $i6$ is a pivot instruction. Therefore, it is
ineffectual. We see that $ILC_{i6} = \{i3,i4,i5\}$. Since $i6$ is the only
dependant of $i5$ ($OLC_{i5} = \{i6\}$), $i5$ is also ineffectual. Similarly, $i3$
is also ineffectual because $OLC_{i3} = \{i5, i6\}$. However, $i4$ is not
ineffectual because there is an instruction ($i14$) in $OLC_{i4}$ which is not
ineffectual.

\begin{definition}[Root Ineffectual Instruction]
An ineffectual instruction is a \textbf{\em Root Ineffectual Instruction} if all
the instructions that it is data dependent on are effectual instructions.
\end{definition}
For example, $i3$ is a root ineffectual instruction assuming that all the
instructions that it is data dependent on are effectual.

\begin{definition}[Ineffectual Graph]
An \textbf{\em Ineffectual Graph} of instructions is a maximal graph, with
ineffectual instructions as nodes, and data dependence relationships as edges.
An ineffectual graph may have one or more roots, and one or more pivots.
\end{definition}
For example, instructions $i3$, $i5$, and $i6$ form an ineffectual graph,
assuming that $i3$ is a root ineffectual instruction, and $i3$ and $i5$ have no
other instructions dependent on them.

In this work, we do not consider memory dependencies (including predicated
\texttt{store} instructions) and stop extending the ineffectual graph when we
encounter a memory operation. The number of ineffectual instructions identified
could increase if we take into account memory dependencies as well, but would
complicate the hardware circuitry further. We will explore memory dependencies
in a future work.

\section{Proposed Architecture}
\label{sec:arch}

\begin{figure*}
\begin{center}
\includegraphics[height=2.25cm]{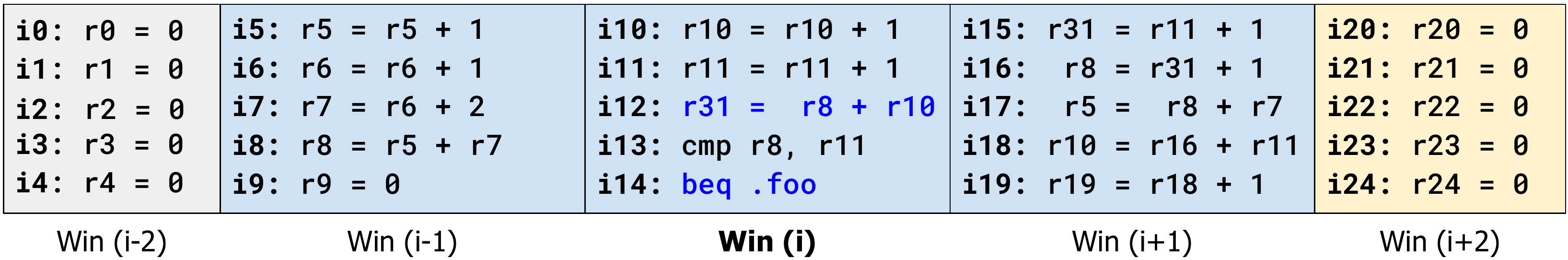}
\caption{A snippet of dynamic micro-ops grouped in windows of 5 instructions
  each.
  \texttt{i14} is marked as pivot because it is an accurately predicted branch
  and \texttt{i12} is marked as pivot because its result is overwritten (by
  \texttt{i15}) before it is used.}
\label{fig:example_code}
\end{center}
\end{figure*}

The modifications can be broadly grouped into four categories: (i) identifying
ineffectual instructions (and tagging them appropriately), (ii) execution of
instructions speculated as ineffectual, (iii) detecting instances where
instructions were misspeculated as ineffectual, and (iv) recovery from such
misspeculations.
For the purpose of discussions, we assume the microarchitecture of the base
system to be similar to Intel's Tiger Lake -- that has a single (unified)
execution cluster that executes all instructions. However, our proposal is
relevant to all modern microarchitectures. We discuss the operation of each of
the above categories in detail below.

We propose to logically organize the dynamic instruction stream into windows of
10 instructions each. Such windowing of instructions helps in identifying
ineffectual instructions as well as in elegant checkpointing and recovery --
the window boundaries form the checkpoints to be restored to in case of a
misspeculation. We use the snippet of 25 (dynamic) instructions shown in the
Figure~\ref{fig:example_code} as an example to discuss the working of the
proposed architecture. Here, we have assumed a windows size of 5 instructions in the
interest of space.

\subsection{Identifying ineffectual instructions}
\label{sec:identification}

\begin{figure*}
\begin{center}
\includegraphics[height=2.25cm]{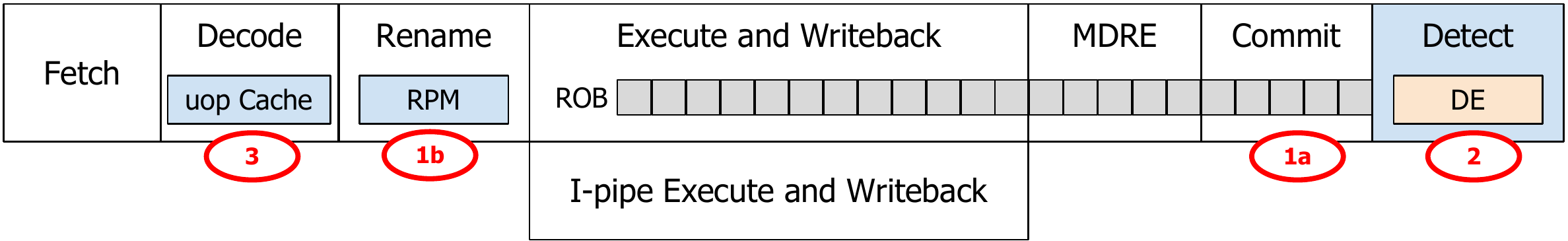}
\caption{Relevant modifications to the pipeline for identifying ineffectual
  instructions. \Circled{1a} Commit stage marks pivots. \Circled{1b} Rename
  stage marks RI\_pivots. \Circled{2} Detect stage identifies ineffectual
  instructions using Algorithm~\ref{alg:identify}. \Circled{3} Ineffectual
  instructions are tagged in the micro-op cache.}
\label{fig:identify}
\end{center}
\end{figure*}

Figure~\ref{fig:identify} shows the three modifications required for identifying
ineffectual instructions: (i) a {\em Register Producer Map} (RPM) is maintained
in the Renamer to identify pivot instructions, (ii) a {\em Detection Buffer}
(DB) which holds recently committed instructions that will be used for
ineffectuality analysis, and (iii) a {\em Detection Engine} that analyzes the
instructions in the DB to identify ineffectual instructions. We discuss each of
them in detail below.

\subsubsection{Identifying Pivots}

The ROB marks the following classes of instructions as pivots at the time of
commit: (i) branch instructions whose outcome was accurately predicted ($i14$ in
the snippet), (ii) predicated instructions whose predicate was accurately
predicted to be false, and (iii) indirect jump instructions whose target was
accurately predicted. The Renamer uses the RPM to identify {\em Register
  Ineffectual pivots} (RI pivot) -- instructions that produce register values
that are not consumed by any other instructions (condition 4 of
Definition~\ref{def:pivot}). The instruction $i12$ in the snippet shown is an
example of such a pivot. Although \texttt{store} instructions could be pivots,
we do not consider them in this work.

Each entry of the RPM corresponds to an architectural register and contains two
fields: (i) a \texttt{has\_dependants} bit that indicates if the results
produced by this instruction is consumed by any other instruction and (ii)
\texttt{producer\_ID} which indicates the window ID of the producer.

The Renamer does the following for every instruction $i$ that is being renamed.
First, the RPM entries corresponding to the source registers of $i$ are looked
up and their respective \texttt{has\_dependants} bit is set to {\tt true}. Next,
if the current value of the destination register of $i$ currently being renamed
has never been consumed and its corresponding producer $p$ occurred within the
current window or the previous window, then we mark that producer $p$ as a RI
pivot by setting the \texttt{RI\_pivot} bit in the ROB entry of $p$ to {\tt
  true}. Consider the renaming of the instruction $i15$ in the example. We see
that the destination register $r7$ would have its \texttt{has\_dependants} bit
set to $false$, and that the corresponding producer ($i12$) occurs in the
previous window. Therefore, we mark the producer instruction $i12$ as a RI
pivot. Lastly, the \texttt{producer\_ID} field of the RPM entry corresponding to
the destination register of $i$ is appropriately set.

It maybe noted that there are cases where an instruction $i$ is one whose result
is never consumed, but the next producer of the destination register of $i$
occurs beyond the next window. In such a case, we conservatively do not mark
instruction $i$ as a RI pivot.

\subsubsection{Identifying ineffectual instructions}
\label{subsec:identification}
The instructions that are committed by the ROB are written into the DB in
program order (similar to \cite{catch}).
The DB is organized as a circular buffer of 40 entries (indexed 0 to 39) and is
treated as 4 windows of 10 entries each. For example, during the processing of
the example code snippet in Figure~\ref{fig:example_code}, the contents of the
oldest window ($i-2$) is overwritten by instructions in the newest window
($i+2$). The ineffectual instruction detection procedure is executed when the
first three windows (30 entries) of the DB are full -- entries $0$ to $29$ are
occupied, with entry $0$ being the oldest instruction.

\begin{algorithm}
\caption{Ineffectual instruction detection procedure}\label{alg:identify}
\begin{algorithmic}[1]
\Procedure{init()}{}
    \For{$i \in [0,29]$}
        \State i.tag = effectual
    \EndFor
\EndProcedure         

\Procedure{analyze\_OLC}{$inst$}
    \ForAll{$s \in $getSucc($inst$)}  
        \If{$s.tag$ == effectual}
            \State \Return false
        \EndIf
    \EndFor
    \ForAll {$o\_inst \in [idx(inst)+1, 29]$}
        \If {dest($inst$) == dest($o\_inst$)}
            \State \Return true
        \EndIf
    \EndFor
    \State \Return false
\EndProcedure

\Procedure{analyze\_ILC}{$inst$}
    \ForAll{$p \in $getPred($inst$)}  
        \If {analyze\_OLC($p$)}
            \State $p.tag$ = ineffectual
            \State analyze\_ILC(p)
        \EndIf
    \EndFor
\EndProcedure   

\Procedure{identify()}{}
    \For{$inst \in [19, 10]$}
        \If{$inst$ is a pivot}
            \State $inst.tag$ = ineffectual
            \State analyze\_ILC($inst$)
        \EndIf
    \EndFor    
\EndProcedure
    
\end{algorithmic}
\end{algorithm}

\begin{figure}
\begin{center}
\includegraphics[width=0.975\columnwidth]{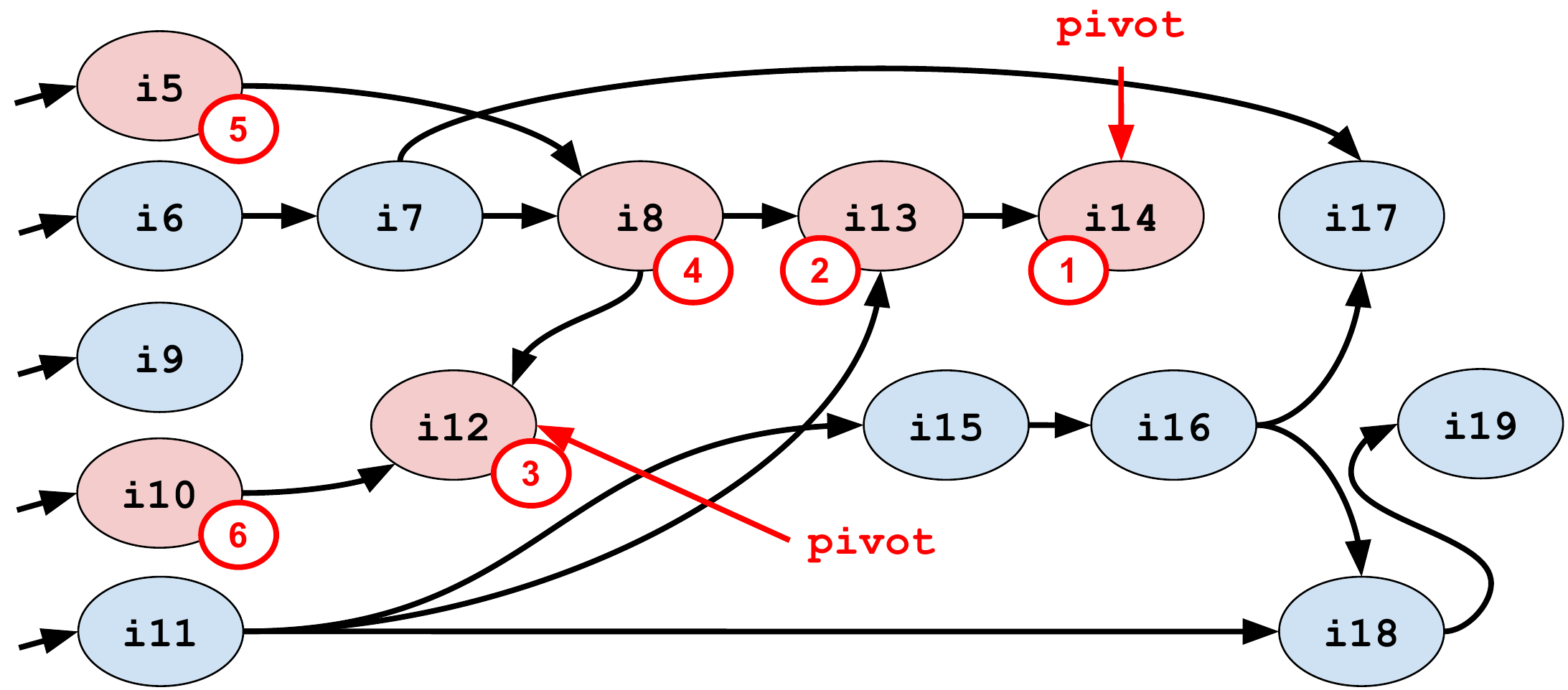}
\caption{Ineffectuality labeling of the dynamic micro-ops shown in
  Figure~\ref{fig:example_code}}
\label{fig:detected_ineffectual}
\end{center}
\end{figure}

The Detection Engine (DE) employs Algorithm~\ref{alg:identify} to identify
ineffectual instructions. Figure~\ref{fig:detected_ineffectual} shows the data
dependencies between the instructions and the order in which
Algorithm~\ref{alg:identify} marks instructions in our example as ineffectual.

The DE starts by looking for pivot instructions in the second window (window
$i$) in the reverse order (entries $19$ to $10$) as shown in the
\texttt{identify()} function. For each instruction in this window that is marked
as pivot (by the ROB), we tag it as ineffectual and we look for the instructions
in its input logic cone (ILC) whose results are solely consumed by ineffectual
instructions and tag them as ineffectual using the \texttt{analyze\_ILC()}
function. For every new instruction that we determine to be ineffectual, we
analyze its \texttt{ILC} recursively to determine more instructions that could
be ineffectual (in line 8).

The \texttt{getPred()} function returns all the instructions {\bf excluding}
memory operations, whose results are directly used by this instruction. The
\texttt{getPred()} function looks for predecessor instructions only within the
DB and therefore, the list of predecessor instructions returned may not be
complete. Therefore, our detection procedure to find ineffectual instructions is
quite conservative. The \texttt{getSucc()} function returns all consumers of an
instruction within the first three windows of the DB. Any successors outside the
first three windows are treated conservatively. The destinations that are
compared in Line 9 are registers only (and not memory locations).

Let us consider the working of this algorithm on our instruction snippet. The
instructions $i12$ and $i14$ are marked as pivots by the ROB to begin with. The
algorithm initializes all the instructions in the windows $i-1$, $i$ and $i+1$
as effectual. Then instruction $i14$ is considered first. Since this is a pivot,
it is first marked as ineffectual \Circled{1}, and its predecessor $i13$ is
analyzed next. Firstly, the only successor of $i13$ in the region under analysis is
$i14$, and the latter has already been tagged as ineffectual. Secondly, $i15$
overwrites the results (flags register) of $i13$, confirming that there are no more
successors of $i13$. Hence, $i13$ is marked as ineffectual \Circled{2} and its
predecessors (\{$i8$,$i11$\}) are analyzed next. The OLC analysis of $i8$ reaches
$i12$ and finds it to be effectual. Therefore, we do not mark $i8$ as
ineffectual. Similarly, the OLC analysis of $i11$ reaches $i15$ and finds it to
be effectual. Therefore, we do not mark $i11$ as ineffectual. The algorithm then
considers the next pivot $i12$ as it continues to scan the second window in the
reverse order. It marks $i12$ as ineffectual \Circled{3} and analyzes its
predecessors (\{$i8$,$i10$\}) next. Let us assume that the algorithm considers
$i8$ first. The OLC analysis of $i8$ now finds all its successors $\{i12,i13\}$
to be ineffectual. $i16$ overwrites the results of $i8$ confirming that the latter
has no other successors. Hence, $i8$ is marked as ineffectual \Circled{4}.
The predecessors of $i8$ (\{$i5$,$i7$\})
are considered next (recursively) although we are yet to analyze $i10$.
The OLC analysis of $i5$ now finds all its successors $\{i8\}$
to be ineffectual. $i17$ overwrites the results of $i5$ confirming that the latter
has no other successors. Hence $i5$ is marked as ineffectual \Circled{5}.
\texttt{getPred($i5$)} returns $\emptyset$ because the window containing its
predecessors have been overwritten (in the DB). The algorithm now backtracks
and comes out of the recursion to analyze $i7$. Since $i17$, a successor of $i7$,
is effectual, $i7$ is deemed effectual. Finally, the algorithm analyzes $i10$
(the predecessor of $i12$, after $i8$) to find no effectual instructions in its
OLC (\{$i12$\}). Hence it marks $i10$ as ineffectual \Circled{6}. The
algorithm terminates when \texttt{getPred($i10$)} returns $\emptyset$.

Once the ineffectual instructions are identified, they are tagged appropriately
in the micro-op cache. There is one additional bit, called the
\texttt{ineffectual} bit, per micro-op (instruction) used for this.
Additionally, we propose to increase the number of micro-ops per way in the
micro-op cache by the width of the secondary pipeline (see
Section~\ref{ssec:execute}) while keeping the overall size of the micro-op cache
unchanged. This reorganization will help feed the wider back-end without
increasing the area overhead. The oldest window is then discarded.

Our empirical studies (see Section~\ref{sec:eval}) show that ineffectuality is a
phenomenon that is quite local to pivot instructions and hence it is rare for an
ineffectual instruction to be found farther than a window away. This justifies
our choice of the window size and DB size, as well as conservatively restricting
our search for predecessors and successors to just three windows of the DB.

Finally, this detection procedure is off the critical path because it is
performed after instructions are committed by the ROB. Hence, its performance
overhead is negligible.

\subsection{Executing ineffectual instructions}
\label{ssec:execute}

\begin{figure*}
\begin{center}
\includegraphics[height=2.3cm]{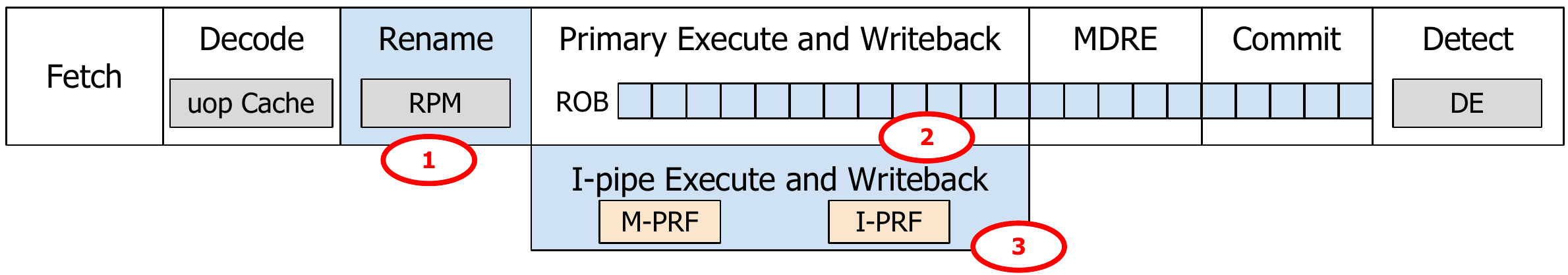}
\caption{Relevant modifications to the pipeline to execute ineffectual
  instructions. \Circled{1} Renamer maps architectural registers to PRF or I-PRF
  entries appropriately. \Circled{2} ROB entries are made even for ineffectual
  instructions to support in-order commit. \Circled{3} Primary pipe executes
  effectual instructions and writes its results to PRF and M-PRF. The I-pipe
  executes ineffectual instructions and writes its results to I-PRF.}
\label{fig:execute}
\end{center}
\end{figure*}

There are three key changes required to the micro-architecture for executing
ineffectual instructions: (i) addition of a new secondary cluster, called {\em
  I-pipe} with additional register files associated with it, (ii) modifications
to the Renamer to support the additional register files as well as increased
front-end bandwidth, (iii) modifications to the ROB to support windowing and
track ineffectual instruction execution.

The I-pipe is a simple, multi-issue, in-order pipeline that executes only
ineffectual instructions. The effectual instructions continue to be executed on
the primary cluster as in the base system. This I-pipe has a separate register
file~\cite{registerwritespecialization} called {\em Ineffectual physical
  register file} (I-PRF) to temporarily store the data generated by ineffectual
instructions. The number of physical registers in the I-PRF is equal to the
number of architectural registers supported by the machine. The I-PRF has two
read ports and one write port. The ineffectual instructions can depend on values
produced by effectual instructions. However, we do not want the ineffectual
instructions to compete for the ports of the PRF. We also wish to avoid complex
by-pass circuitry from the primary cluster to the I-pipe. Hence, we propose to
keep a copy of the PRF, called {\em Mirrored-PRF} (M-PRF) (similar to
\cite{kessler1999alpha}). All the writes to the PRF are forwarded to the
secondary cluster to be written into the M-PRF. The M-PRF has the same number of
write ports as the PRF, but has only two read ports. An ineffectual instruction
that depends on the results produced by an effectual instruction reads the M-PRF
now instead of the PRF. The reduced number of read ports in the M-PRF results in
a simpler structure as compared to the PRF. The combination of I-PRF and M-PRF
ensures that the execution units (EUs) in the secondary cluster would never have
to fetch data from the primary cluster. Also, by definition
(Definition~\ref{def:ineffectual}), no effectual instruction would ever require
data produced in the secondary cluster. Therefore, the execution of effectual
instructions never waits for ineffectual instructions to complete. The execution
of ineffectual instructions is only to verify the ineffectuality speculation.

The Renamer now maps the architectural registers to the appropriate physical
register in the PRF/I-PRF. The width of the Renamer by the width of the I-pipe
to sustain the increased throughput of the system.

Finally, we increase the number of ROB entries to $370$ (up from $352$), with
window IDs ranging from $0$ to $36$, because it is a multiple of the window size
and includes one additional window for recovery (as discussed later). We also
add one bit, called \texttt{ineffectual\_ROB\_entry} to each entry of the ROB to
record if the instruction is ineffectual or not. This is in addition to the
\texttt{RI\_pivot} bit that was discussed earlier.

\subsubsection{Instruction flow through execution clusters}
Figure~\ref{fig:execute} shows the execution of instructions in the proposed
architecture. The Renamer could get effectual instructions from either the
decode stage or the micro-op cache, but ineffectual instructions would come only
from the micro-op cache. The number of effectual instructions that the Renamer
renames in each cycle continues to remain the same. If the instruction being
renamed is effectual, the Renamer maps the destination register to the PRF and
marks the \texttt{ineffectual\_ROB\_entry} bit in the ROB entry corresponding to
this instructions to \texttt{false}. However, if the instruction being renamed
is ineffectual, then the Renamer maps the destination register to the I-PRF. The
ineffectual instruction continues to have an entry in the ROB (to maintain
program order), but the \texttt{ineffectual\_ROB\_entry} bit is set to
\texttt{true}. Finally, the Renamer makes an entry for an ineffectual
instruction in the reservation station of the I-pipe (I-RS).

The oldest entry in the I-RS is considered for execution. Its operands are
sourced from either the I-PRF (if its producer was another ineffectual micro-op)
or from the M-PRF (if its producer was an effectual micro-op). Once the operands
are available, the micro-op is executed on functional units that are private to
the I-pipe. There is no complex bypass logic serving ineffectual micro-ops. Once
the execution is complete, the results are written to the I-PRF immediately and
the ROB is notified of the execution completion. The execution of effectual
instructions in the primary cluster is similar to that of the base system except
that it skips the execution of ineffectual instructions (whose
\texttt{ineffectual\_ROB\_entry} is set to \texttt{true}).

The memory subsystem of the proposed architecture is the same as the original
architecture because memory operations are never marked ineffectual.

\subsubsection{Handling I-pipe bottlenecks}
There are rare instances where the number of ineffectual instructions offloaded
to the I-pipe for execution overwhelms the I-pipe. This results in the I-pipe
becoming the bottleneck which further leads to an overall slowdown in the
execution of the application. When the I-pipe becomes the bottleneck, the
Renamer finds the I-RS to be full quite often. We flush the pipeline and reset
the \texttt{ineffectuality} bits in the micro-op cache in such a scenario. This
prevents any instruction from being offloaded into the I-pipe for execution
until the DE identifies and tags instructions in the micro-op cache again.

\subsection{Detecting and recovering from misspeculated ineffectual
  instructions}
\label{ssec:misspec}

\begin{figure*}
\begin{center}
\includegraphics[height=2.3cm]{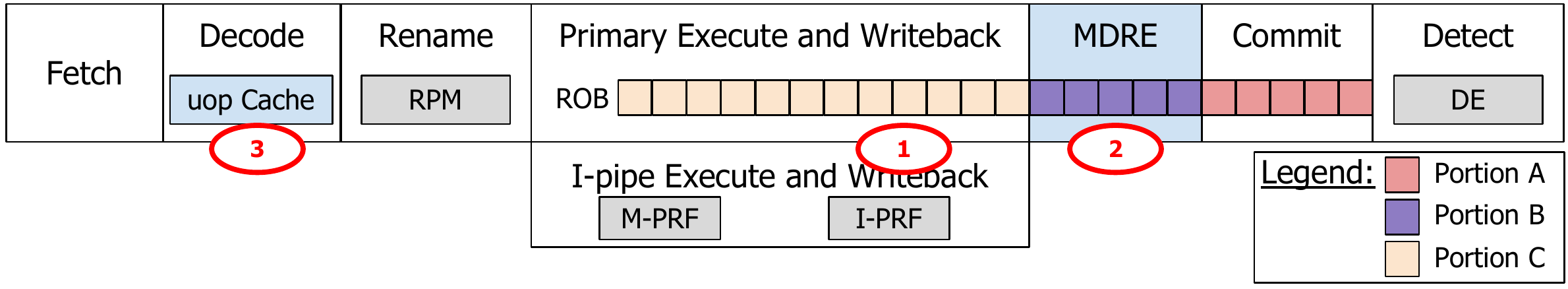}
\caption{Relevant modifications to the pipeline to detect misspeculated
  ineffectual instructions and recover the pipeline. \Circled{1} ROB is
  partitioned into portions A, B, and C. \Circled{2} A new pipeline stage, {\em
    MDRE} is added to detect misspeculations and initiate recovery. \Circled{3}
  The ineffectuality tags in the micro-op cache are reset upon detection of a
  misspeculation.}
\label{fig:detect}
\end{center}
\end{figure*}

There are two kinds of misspeculations possible:
\begin{enumerate}
\item {\em Type-A}: (i) a conditional branch that was deemed ineffectual turned
out to be a mispredicted branch, (ii) a predicated instruction deemed
ineffectual needed to be executed, or (iii) an indirect jump whose target
address was mispredicted.
\item {\em Type-B}: the register result of an instruction deemed ineffectual is
consumed by an effectual instruction. Such misspeculations are an implication of
a preceding Type-A misspeculation.
\end{enumerate}

Figure~\ref{fig:detect} shows the changes to the base system required to detect
instances of misspeculations mentioned above: (i) the ROB is logically grouped
into Portion A (the oldest window), Portion B (the next oldest window), and
Portion C (rest of the windows), and (ii) a {\em Misspeculation Detection and
  Recovery Engine} (MDRE) is added as an additional pipeline stage to detect and
recover from misspeculations.
The MDRE verifies that there is no misspeculation of ineffectual instructions in
the Portion B. If this is indeed the case, all the instructions in the Portion A
can be committed (removed from ROB and inserted into the DB).

The MDRE detects misspeculations in Portion B, only when all the instructions in
it have completed their execution (either in the primary pipe or in the I-pipe).
The MDRE detects Type-A misspeculations when the I-pipe executes the
corresponding control instruction and finds that it was mispredicted. Since
Type-B misspeculations are always in the shadow of an earlier Type-A
misspeculation, we do not expend additional circuitry to detect such
misspeculations. Instead, we employ a conservative recovery mechanism that
handles Type-A misspeculation and all the associated Type-B misspeculations.

Once the MDRE detects a misspeculation, it initiates the rollback of the
pipeline to a valid checkpoint. The checkpoints to rollback to for different
misspeculation types are as follows. For Type-A misspeculations that are a
result of a mispredicted branch instruction or indirect jump instruction, the
first instruction on the correct path is the point from which the pipeline
should resume its execution. Similarly, if the misspeculation is due to
misprediction of the condition of a predicated instruction, then the pipeline
should resume from that predicated instruction. For Type-B misspeculations, the
pipeline should resume its execution from the earliest Root Ineffectual
instruction of the Ineffectual Graph that contained the misspeculated
instruction. We know for sure that all Root Ineffectual instructions will lie
within the current window or the window prior to the current one because the ILC
was not grown over earlier windows (\texttt{getPred()} in
Algorithm~\ref{alg:identify} returned predecessors that are in the current or
the previous window only), and hence we can conservatively roll back to the
start of the previous window. To keep the circuitry simple, we adopt a
conservative strategy of rolling back to the start of Portion A of the ROB for
both misspeculation types. As an example, if instructions $i12$ or $i14$ were
misspeculated (detected when these instruction are in Portion B), we rollback to
$i5$ (start of Portion A during such detection). 
To support such rollbacks, the Branch Order Buffer (BOB) maintains checkpoints
at the start of each window.

In the event of a rollback, the \texttt{ineffectuality} bit (in the micro-op
cache) corresponding to the instructions in Portions A and B are reset to
\texttt{false}. The structures associated with the execution of an instruction
(effectual/ineffectual) such as ROB, RPM, and I-RS are flushed.

\subsection{Committing of instructions}
\label{sec:commit}
When the Commit stage of the system receives the signal to commit from MDRE, it
proceeds to commit the Portion A of the ROB. This requires the commit width to
be the same as the window size. Once done, it pauses and waits for the next
signal from MDRE. All committed instructions are then added to the DB (with
pivot instructions suitably marked). The branch predictor is trained for both
effectual and ineffectual instructions.

\section{Evaluation}
\label{sec:eval}

\subsection{Setup}
We used the Tejas architectural simulator~\cite{tejas} to evaluate our proposed
architecture. The simulation configurations were tuned to resemble the Intel
Tiger Lake architecture~\cite{spectre, tigerlake}. We implemented stride and
stream prefetchers into all caches (including the L1 i-cache). We integrated the
publicly available TAGE-SC-L implementation~\cite{tage_sc_l} into the simulator.
Table~\ref{tab:sim_params} gives the details of the simulation parameters used
for evaluating the proposal.

\begin{table}
\caption{Simulation Parameters}
\label{tab:sim_params}
\footnotesize 
\bgroup
\def\arraystretch{1.15}
\begin{tabular}{|lll|}
  \hline
  \multicolumn{1}{|l|}{\textbf{Parameter}}
  & \multicolumn{1}{l|}{\textbf{Baseline}}
  & \textbf{Proposed} \\
  \hline
  \multicolumn{3}{|c|}{Pipeline} \\
  \hline
  \multicolumn{1}{|l|}{micro-op cache}
  & \multicolumn{1}{l|}{\begin{tabular}[c]{@{}l@{}}2304 entries,\\
                          6 uops/way\end{tabular}}
  & \begin{tabular}[c]{@{}l@{}}2304 entries, \\
      (6 + {\it I-pipe\_width}) uops/way\end{tabular}
  \\ \hline 
  \multicolumn{1}{|l|}{Rename Width}
  & \multicolumn{1}{l|}{6}
  & \begin{tabular}[c]{@{}l@{}}$\le$ 6 effectual \\
      $\le$ {\it I-pipe\_width} ineffectual\end{tabular}
  \\ \hline 
  \multicolumn{1}{|l|}{Register Files}
  & \multicolumn{1}{l|}{280 int, 224 vec}
  & \begin{tabular}[c]{@{}l@{}}PRF: 280 int, 224 vec\\ I-PRF: 16 int, 32 vec\\ M-PRF: 280 int, 224 vec\end{tabular}
  \\ \hline 
  \multicolumn{1}{|l|}{ROB}
  & \multicolumn{1}{l|}{352-entry}
  & 370-entry
  \\ \hline 
  \multicolumn{1}{|l|}{Scheduler}
  & \multicolumn{1}{l|}{160-entry}
  & \begin{tabular}[c]{@{}l@{}}160-entry (in Primary pipe)\\
      {\it I-RS\_size} (in I-pipe) \end{tabular}
  \\ \hline 
  \multicolumn{1}{|l|}{Issue Width}
  & \multicolumn{1}{l|}{10}
  & \begin{tabular}[c]{@{}l@{}}10 (in Primary)\\ {\it I-pipe\_width} (in I-pipe)\end{tabular}
  \\ \hline
  \multicolumn{1}{|l|}{Commit Width}
  & \multicolumn{1}{l|}{6}
  & {\it Window\_size}
  \\ \hline
  \multicolumn{1}{|l|}{Detection Buffer}
  & \multicolumn{1}{l|}{}
  & (4 * {\it Window\_size})
  \\ \hline
  \multicolumn{3}{|c|}{Memory}
  \\ \hline
  \multicolumn{1}{|l|}{LSU}
  & \multicolumn{2}{l|}{128 load entries, 72 store entries,}
  \\
  \multicolumn{1}{|l|}{}
  & \multicolumn{2}{l|}{2 loads and 2 stores per cycle}
  \\
  \multicolumn{1}{|l|}{L1 i-cache}
  & \multicolumn{2}{l|}{32KB, 8-way associative, 4c latency}
  \\
  \multicolumn{1}{|l|}{L1 d-cache}
  & \multicolumn{2}{l|}{48KB, 12-way associative, 5c latency}
  \\
  \multicolumn{1}{|l|}{L2 cache}
  & \multicolumn{2}{l|}{1280KB, 20-way associative, 14c latency}
  \\
  \multicolumn{1}{|l|}{L3 cache}
  & \multicolumn{2}{l|}{3MB per core, 12-way associative, 45c latency}
  \\
  \multicolumn{1}{|l|}{Main memory}
  & \multicolumn{2}{l|}{132c latency}
  \\ \hline
\end{tabular}
** {\it I-pipe\_width}: 2/4/8, {\it I-RS\_size}: 64/128/256, {\it Window\_size}: 10
\egroup
\end{table}

We simulated 19 out of the 23 benchmarks of the SPEC CPU2017 (SPEC2017)
benchmark suite~\cite{spec2017} and 5 out of the 6 benchmarks of
GAPBS~\cite{gaps}. The benchmarks were compiled and statically linked. For the
SPEC2017 suite, we used the $ref$ inputs. We determined the representative
portion of each benchmark of both the suites using
SimPoint~\cite{sherwood2002automatically} and simulated 100 million instructions
from it.

We individually study the impact of constructing the ILCs from different pivot
types. The pivots corresponding to the first three conditions mentioned in
Definition~\ref{def:pivot} are grouped together and is denoted as {\em C} (or
control-only) pivots. The pivots corresponding to the fourth condition is
denoted as {\em D} (or data-only) pivots. We also consider the combination of
the two pivot types and denote it as {\em CD} (or control+data).

\subsection{Results}

\subsubsection{Performance Characterization}
\begin{figure*}
\begin{center}
  \includegraphics[width=0.95\textwidth]{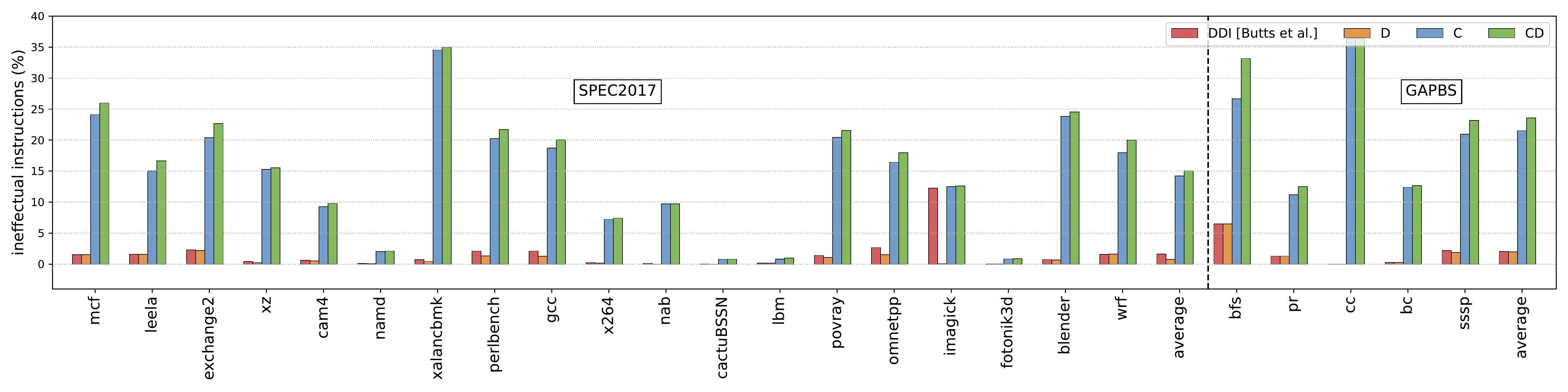}
  \caption{Number of ineffectual instructions identified}
  \label{fig:ineffectual}
\end{center}
\end{figure*}

\begin{figure*}
\begin{center}
\includegraphics[width=0.95\textwidth]{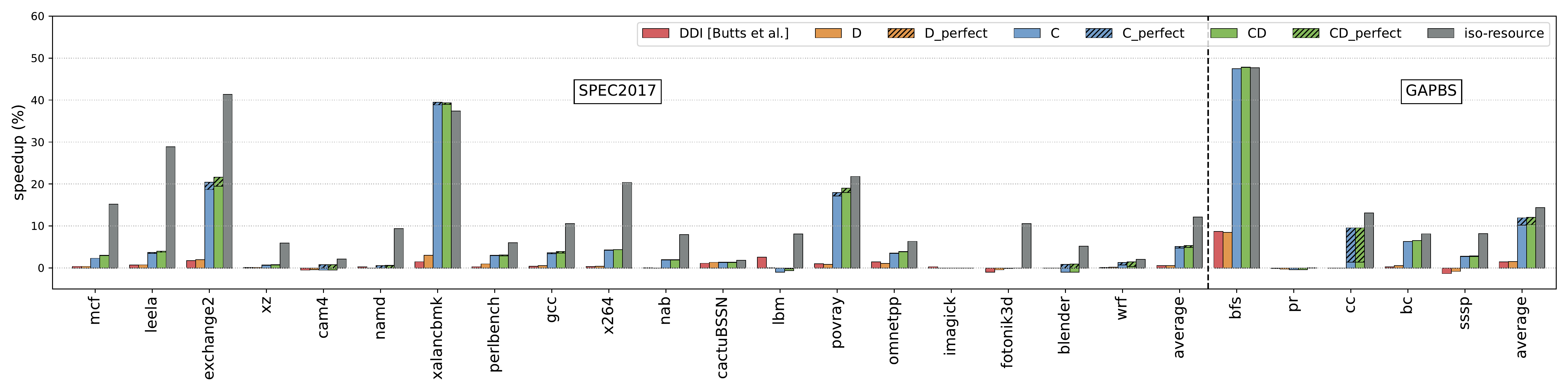}
\caption{Performance gain}
\label{fig:perf}
\end{center}
\end{figure*}

Figures~\ref{fig:ineffectual}~and~\ref{fig:perf} show the number of ineffectual
instructions identified by our proposed scheme and the resulting performance
improvements respectively. We have used an I-pipe of width $4$ (with $128$ I-RS
entries) for this study. The performance gains for other I-pipe widths and
varying I-RS entries is discussed later in this section (see
Figure~\ref{fig:ipipe}).

The ineffectual instructions borne out of the {\em D} pivot type is conceptually
similar to the Dynamically Dead Instructions (DDI) identified by an earlier
work~\cite{butts2002dynamic}. We notice that the number of ineffectual
instructions identified by our scheme under the {\em D} pivot type is similar to
DDI (lesser by $<1$\% on average). This is because of two counteracting
attributes. First, DDI considers \texttt{load} instructions as candidate dead
instructions whereas {\em D} pivot type does not (see {\em imagick} in
Figure~\ref{fig:ineffectual}). Second, our proposed scheme takes into
transitivity of ineffectuality -- instructions whose all consumers are
ineffectual are also ineffectual, whereas DDI does not. We observe that the
performance uplift obtained by both the techniques are similar.

The ineffectual instructions borne out of the {\em C} pivot type are higher in
number, $14.23$\% and $21.53$\% on average, and up to $34.53$\% (seen in {\em
  xalancbmk}) and $36.43$\% (seen in {\em cc}) in the case of SPEC2017 and GAPBS
respectively. The performance uplifts obtained by executing the ineffectual
instructions on I-pipe is $4.74$\% and $10.21$\% on average, and up to $38.85$\%
(seen in {\em xalancbmk}) and $47.44$\% (seen in {\em bfs}) in SPEC2017 and
GAPBS respectively. This is significantly higher than the respective numbers
observed in the case of DDI~\cite{butts2002dynamic} and the {\em D} pivot type.

The {\em CD} pivot type further increases the number of ineffectual instructions
identified and the performance uplifts seen. The number of ineffectual
instructions identified is $15.1$\% and $23.6$\% on average, and up to $35$\%
(seen in {\em xalancbmk}) and $36.43$\% (seen in {\em cc}) in the case of
SPEC2017 and GAPBS respectively. The corresponding performance uplift is
$4.93$\% and $10.31$\% on average, and up to $38.94$ (seen in {\em xalancbmk})
and $47.73$\% (seen in {\em bfs}) in the case of SPEC2017 and GAPBS
respectively.

We observe a trend where the performance gain is higher if the number of
ineffectual instructions identified is also high. The offloading of the
execution of ineffectual instructions to the I-pipe results in dedicating the
resources of the primary pipe, which is a complex out-of-order cluster,
exclusively for the execution of effectual instructions. However, the increase
in performance cannot be directly equated to the number of ineffectual
instructions offloaded. The performance of the application could depend on other
factors such as cache behavior and degree of Instruction-level Parallelism among
effectual instructions which is beyond the purview of the proposed scheme. This
effect is observed in {\em mcf} where the number of ineffectual instructions
identified is high ($26$\%), but performance gain is modest ($3$\%).

We notice benchmarks such as {\em cactuBSSN}, {\em lbm}, and {\em fotonik3d}
that do not have too many ineffectual instructions. Consequently, we observe
meagre gains ($1.34$\%, $-0.17$\%, and $-0.05$\% respectively) from our proposed
techniques in these benchmarks. We also notice benchmarks such as {\em blender}
where the number of ineffectual instructions identified is high ($24.5$\%) but
the performance drops by $1$\%. This is because of the I-pipe becoming the
bottleneck due to a large number of data dependencies between ineffectual
instructions which results in a low issue-rate in the I-pipe.

\subsubsection{I-pipe Characterization}

\begin{figure}
\begin{center}
\includegraphics[width=0.9\columnwidth]{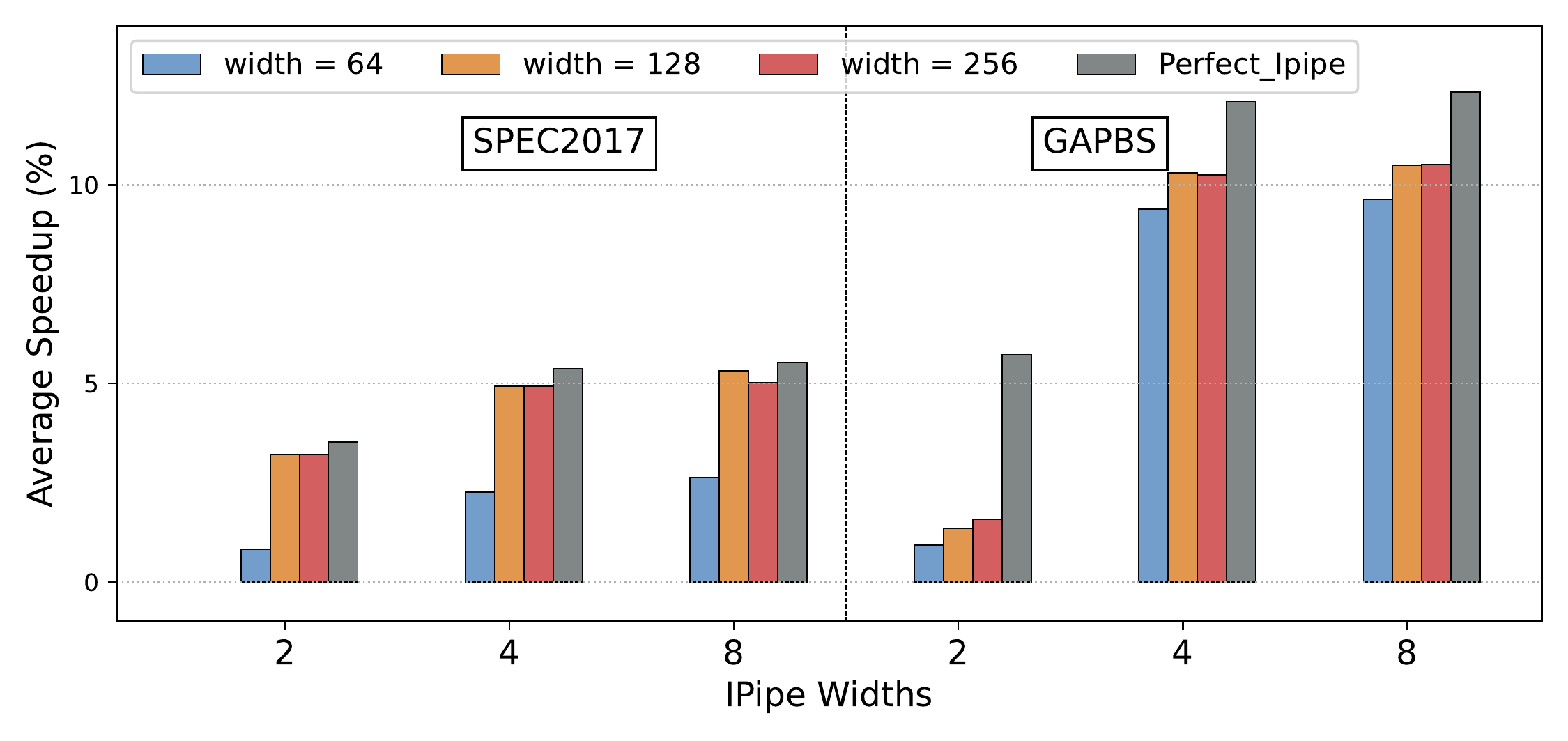}
\caption{Average speedup for SPEC2017 and GAPS suites for varying I-pipe parameters}
\label{fig:ipipe}
\end{center}
\end{figure}

Given the large number of ineffectual instructions identified by the proposed
technique, the I-pipe should be designed in a way that it does not become a
bottleneck. We studied the performance of the I-pipe by setting the number of
I-RS entries ({\it I-RS\_size}) to $64$, $128$ and $256$. For each of these
values, we simulated issue widths ({\it I-pipe\_width}) of $2$, $4$ and $8$. We
also simulated perfect I-pipes ({\it Perfect\_Ipipe}) -- a perfect I-pipe of
width {\it I-pipe\_width} issues {\it I-pipe\_width} ineffectual instructions in
every cycle regardless of data and structural hazards. The perfect I-pipe is a
hypothetical reference that gives an estimate of the potential performance gain
possible by offloading the execution of ineffectual instructions to an I-pipe of
the corresponding width.

The hatches on the bars of Figure~\ref{fig:perf} indicate the performance
uplifts obtained by the perfect I-pipe for each benchmark. We notice that the
performance gains of the proposed I-pipe design is only marginally lesser than
that observed with the perfect I-pipe (difference of $0.4$\% and $1.75$\% on
average in the case of SPEC2017 and GAPBS respectively). This justifies our
choice of a simple, energy-efficient, multi-issue in-order I-pipe. The only
cases where the difference in performance is large are {\em blender} (in
SPEC2017) and {\em cc} (in GAPBS).

Figure~\ref{fig:ipipe} shows the average performance gain across SPEC2017 and
GAPBS for the different I-pipe parameters (under the {\em CD} pivot type). We
observe that when the issue width increases from $2$ to $4$, the average
performance gain improves from $3.2$\% to $4.93$\% as seen in the case of
SPEC2017 with {\it I-RS\_size} as $128$. However, on further increasing the
width to $8$, we only observe an modest increase (to $5.32$\%) in average
performance gain. We observe similar trends in the case of GAPBS as well. We
perform similar analysis by varying the number of I-RS entries. We observe that
when the number of I-RS entries ({\it I-RS\_size}) increases from $64$ to $128$,
the average performance gain improves from $2.26$\% to $4.93$\% as seen in the
case of SPEC2017 with {\it I-pipe\_width} as $4$. However, on further increasing
the number of entries to $256$, we do not observe any increase in average
performance gains. We observe similar trends in the case of GAPBS. This is the
reason we chose ({\it I-RS\_size}$=128$ and {\it I-pipe\_width}$=4$) as the
primary design point for the performance studies.

\subsubsection{Iso-resource studies}
We also compare the performance gains obtained by the proposed technique to a
hypothetical out-of-order pipeline, labelled {\em iso-resource} in
Figure~\ref{fig:perf}, that consumes similar resources ($D10\_R10\_I14\_C10$
with larger instruction window ($160+128=288$) and PRF entries ($280+16=296$
int, $224+32=256$ vec)). We observe that the average performance gain of the
hypothetical pipeline ({\em iso-resource}) is $12.13$\% and $14.37$\% in the
case of SPEC2017 and GAPBS as compared to the corresponding $4.93$\% and
$10.31$\% as seen in our proposed scheme. The {\em iso-resource} pipeline
however suffers from the second, third and the fourth caveats mentioned in
Section~\ref{sec:intro}, but our proposal does not.

\subsubsection{Misspeculation Characterization}

\begin{figure*}
\begin{center}
\includegraphics[width=0.95\textwidth]{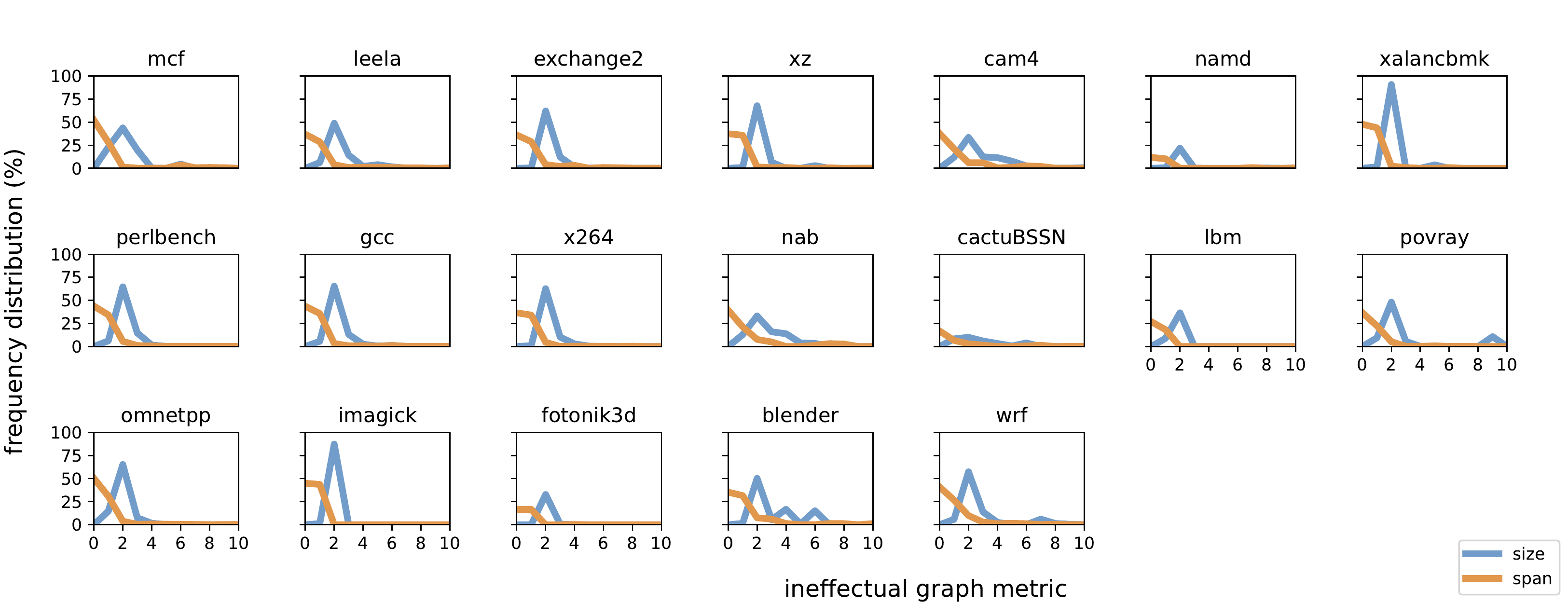}
\caption{Size and span distribution of ineffectual graphs}
\label{fig:size_span}
\end{center}
\end{figure*}

\begin{table}
\begin{center}
\caption{Average MPKI - SPEC2017}
\label{tab:mpki}
\footnotesize
\bgroup
\def\arraystretch{1.15}
\begin{tabular}{|l|c|c|c|c|}
  \hline
  CD & Branch & Predicate & Indirect Jump & Total \\
  \hline
  SPEC2017 & 3.12 & 0.08 & 1.38 & 4.58 \\
  GAPBS    & 6.01 & 0.01 & 0.11 & 6.13 \\
  \hline
\end{tabular}
\egroup
\end{center}
\end{table}

Table~\ref{tab:mpki} shows the number of Type-A misspeculations per
kilo-instruction (mpki) for different misspeculation types. The Type-A
misspeculation are further divided into the constituent control-flow
misspeculations. We see that the total number Type-A misspeculations is quite
less, only $4.58$ and $6.13$ (mpki) on average in the case of SPEC2017 and GAPBS
respectively. Such infrequent Type-A misspeculation motivates us to adopt
conservative flushing of the entire ROB on control-flow misspeculations. This
strategy adds only a marginal overhead to the otherwise partial flush of the ROB
to recover from the control-flow misspeculation (in the baseline). The Type-B
misspeculations are a consequence of an earlier Type-A misspeculation and hence
is not shown separately.

\subsubsection{Sensitivity to Branch Predictors}

\begin{table}
\begin{center}
\caption{Sensitivity of Performance Gain to Branch Predictor Accuracies. }
\label{tab:bpred}
\footnotesize
\bgroup
\def\arraystretch{1.15}
\begin{tabular}{|l|l|c|c|c|c|}
  \hline
  \multirow{4}{*}{}& \multirow{4}{*}{Predictor} & Avg.  & Avg. & Avg. \\
                   & & pred. & Insn.  & Perf. \\
                   & & accuracy  &  dropped & gain  \\
                   & &  (\%) &  (\%) & (\%) \\
  \hline
  \multirow{2}{*}{SPEC2017} & TAGE      & 91.70  & 14.21 & 3.91 \\
                            & TAGE-SC-L & 95.75  & 15.1  & 4.93 \\
  \hline
  \multirow{2}{*}{GAPBS} & TAGE      & 86.78  & 22.54  & 10.45 \\
                         & TAGE-SC-L & 92.19  & 23.6   & 10.31 \\
  \hline
\end{tabular}
\egroup
\end{center}
\end{table}

The number of ineffectual instructions identified is dominated by the accurate
prediction of the control flow. Therefore, the accuracy of the branch predictor
plays a crucial role in the efficacy of this technique. Table~\ref{tab:bpred}
shows the performance gains achieved when TAGE~\cite{tage} and
TAGE-SC-L~\cite{tage_sc_l} branch predictors are used. We observe that the
higher prediction accuracies achieved by TAGE-SC-L help identify more
ineffectual instructions. When a baseline system that uses a TAGE predictor is
augmented with our proposed scheme, the average performance gain is $3.91$\% and
$10.45$\% in the case of SPEC2017 and GAPBS respectively. The same in the case
of a base system that uses a TAGE-SC-L predictor is $4.93$\% and $10.31$\%
respectively. This shows that significant performance gains are possible when
our proposed technique is used with state-of-the-art predictors.

\subsection{Ineffectual graph characterization}

We now consider the characteristics of the ineffectual graph constructed by the
DE during the execution of the workload. For this experiment, we consider a
window size of 80 instructions so that the ILC and OLC constructions are less
restricted at the window boundaries.

The {\em size} of an ineffectual graph is defined as the number of ineffectual
instructions (nodes) in the graph. We observe that the graph sizes are mostly
around $2$ -- $4$ instructions, and are rarely over $8$. The {\em span} of an
ineffectual graph is defined as the number of instructions between the earliest
root in the graph and the latest pivot (in program order), including effectual
instructions. Figure~\ref{fig:size_span} shows the frequency distribution of the
size and the span of the ineffectual graph (under the {\em CD} pivot type). We
observe that the span is usually under $4$, and rarely over $8$, This indicates
that the ineffectual instructions are usually in the vicinity of pivots and
hence led us to fix the window sizes in our proposal to $10$.

\subsection{Memory requirements}
The M-PRF is the largest structure that needs to be added to the base system.
This requires an additional space of $17 KB$. Similarly, the I-PRF
requires $2 KB$ of additional space. The rest of the structures put together
such as DB, I-RS, RPM, additional bits in the micro-op cache, and additional ROB
entries require $< 5 KB$. Therefore, the proposal requires an
additional $24 KB$ of memory.

\section{Conclusions}
\label{sec:conc}

In this work, we propose a technique to detect ineffectual instructions --
instructions that do not have any effect on the program output, in a dynamic
instruction stream. Our proposed scheme identified a significant number (up to
$35$\% and $36.4$\% in SPEC2017 and GAPBS respectively) of ineffectual
instructions. We then proposed an architecture that steers ineffectual
instructions to a dedicated, secondary cluster for its execution. This allowed
the resources of the primary cluster to be reserved for effectual instructions
only. Such ineffectuality-based clustering resulted in the time-consuming,
inter-cluster communication being used by delay-tolerant, ineffectual
instructions only. The uncontended execution of effectual instructions led to
performance improvements of up to $38.9\%$ and $47.7\%$ (and an average gain
of $4.9\%$ and $10.3\%$) in the case of SPEC2017 and GAPBS respectively, as
compared to state-of-the-art processors.

\bibliographystyle{IEEEtranS}
\bibliography{refs}

\end{document}